# A New Trend of Pseudo Random Number Generation using QKD

Omer K. Jasim Mohammad
Ph.D. Student – Computer Science
Al-Ma'arif University College
Anbar-Iraq

Safia Abbas
Computer Science Instructors
FCIS-Ain Shams University
Cairo-Egypt

El-Sayed M. El-Horbaty
Head of Computer Science
FCIS-Ain Shams University
Cairo-Egypt

Abdel-Badeeh M. Salem
Head of B.M. Info. &Know. Eng.
FCIS-Ain Shams University
Cairo-Egypt

## ABSTRACT
Random Numbers determine the security level of cryptographic applications as they are used to generate padding schemes in the encryption/decryption process as well as used to generate cryptographic keys. This paper utilizes the QKD to generate a random quantum bit rely on BB84 protocol, using the NIST and DIEHARD randomness test algorithms to test and evaluate the randomness rates for key generation. The results show that the bits generated using QKD are truly random, which in turn, overcomes the distance limitation (associated with QKD) issue, its well-known challenges with the sending/ receiving data process between different communication parties.

## 1. INTRODUCTION
Random numbers have many uses in cryptography such as key streams of one-time pads, secret keys of symmetric cipher systems, public key parameters, session keys, nonce, initialization vectors and salts and random numbers sequences are of crucial importance in almost every aspect of modern digital cryptography, having a significant impact on the strength of cryptographic primitives in securing secret information by rendering it unknown, un-guessable, unpredictable, and irreproducible for an adversary [1, 2, 3].

Random number generators (RNG) are composed of two main families, (i) the True Random Number Generators (TRNGs) which is exploit the physical phenomena that contain a part of incertitude, and (ii) Pseudo Random Number Generators (PRNGs) are based on deterministic algorithm and it considered more appropriate for security applications [3, 4,5].

The aim of RNG is to produce numbers independent and intricately distributed. A common and easy way to generate random numbers relies on cryptographically secure PRNG [15].

In this paper, the exact QKD protocol (BB84) used to generate truly random and secure bit sequence as a cryptographic key. BB84 protocol is considered a famous protocol in QKD scheme, with this protocol several random bits are required to generate a single shared random bit in which a single photon is sent and received [16, 17].

Moreover, the randomness of quantum keys is investigated to study and testing, these two operations are implementing on the bits generated from QKD using BB84 protocol. In turn on, solving the limited distance coverage, its famous problem associated with quantum key exchanging [12, 14, 17].

To achieve the truly testing, NIST [15], TestU01 [16], and DIEHARD [3] are implanted to test the quantum bits randomness.

The rest of the paper is organized as follows: Section 2 explains the famous articles and algorithms using to generate truly random numbers. The QKD system and the BB84 protocol phases are given in Section 3. Section 4 explains the methods for random testing. Section 5 shows the simulation and implementation of BB84-QKD protocol. The testing and the analysis are given in Section 6. Section 7 presents the conclusions and future works.

## 2. RELATED WORKS
Recently, several implementations of RNGs were realized, almost of them focused on the timing generation or what is the type of operations are based to generate the random number.

For example, Mathilde et al. [1] focus on the generation method, by describing a special experiment by embedded a generator into a processor. Then, the NIST is used as a statistical tool for testing the number of generations. Since the efficiency of such experiment depends on the efficiency of the process, the truly bit number generation is impossible to be achieved.

In [7, 8], the Authors utilizes DS-CDMA communication systems to generate randomness depending on Chaotic maps. Despite their success in the random generation process, the verification of truly randomness is missing. So the proposed system cannot be guaranteed to be used by the encryption algorithms.

In [4, 5], the authors implement the Fibonacci and Gaussian as a mathematical method for the random generation process. They test and compare the results from both methods based on the NIST algorithm. The test results show that the random generation using Fibonacci is more efficient than using Gaussian.

Christian et al. [9] Propose a new coherent state QKD protocol that does away with the need to randomly switch between measurement bases. This protocol provides significantly higher random secret key rate with increased bandwidths than another protocol. At this point the authors did not specify the phase which adopted to generate a random number.

In [6, 7] the authors prove the full randomness can indeed be certified using quantum non-locality under the minimum possible assumptions, the results they represent a special quantum protocol for full randomness amplification, finally, they open a new path for device-independent protocols under minimal assumptions. This study is considered a weakness, because it's not providing a practical model or simulation model.

Therefore, we can conclude that the randomness is an essential field of the security applications, especially for cryptographic system, digital signature, and authentication.

## 3. QKD SYSTEM

The QKD enables secret quantum keys exchanging between two different parties through a communication channel, like optical fiber. It is an alternative to the classical encryption techniques; symmetric or asymmetric, and used to solve the common current scheme problems, such as key distribution management, availability and attack defeating [3, 17, and 18]. In order to exchange the keys between parts, QKD uses the BB84 protocol for single photon polarization over two channels (quantum and classical) see Figure 1. However, both sender and receiver must use devices that generate and detect light pulses with dissimilar polarizations and implement four basic phases of BB84 protocol [11].

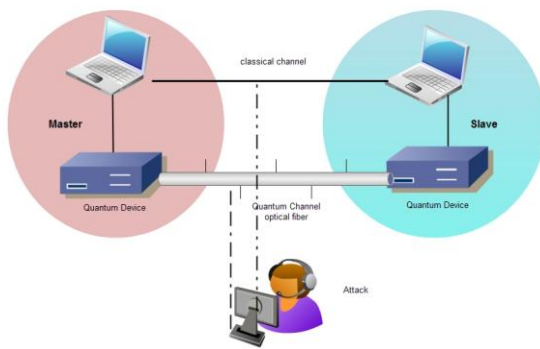

**Fig 1: QKD components**

- Raw Key Extraction (RK)

The main purpose of raw key extraction phase is to eliminate all possible errors occurred during the bits discussion (generation and transmission) over quantum channel. Negotiated parties (i.e. Sender and receiver) compare their filter types used for each photon, unmatched polarization is eliminated otherwise, bits are considered [10, 11].

- Error Estimation (EE)

The negotiation process might occur over a noisy quantum and unsecured (public) classical channel. Such channel can cause a partial key damage or physical noise of transmission medium [9, 13]. To avoid such problems, both sides determine an error threshold value "Emax" when they are sure that there is no eavesdropping on a transmission medium. So as to calculate the error percentage (E), the raw bits are monitored and compared with the Emax, if E > Emax, then it is probably either unexpected noise or eavesdroppers.

- Key Reconciliation (KR)

Key reconciliation is implemented to minimize the mentioned errors associated with the key as much as possible. It divides the raw key into blocks of K bits, then parity bit calculations is done for each block [10, 11 for more details]. Both blocks and bit calculations are performed for N-rounds that depend on the length of the raw key, where the value of N is completely negotiated by different parties.

- Privacy Amplification (PA)

Privacy Amplification is the final step in the quantum key extraction. It is applied to minimize the number of bits based on equation 1 that an eavesdropper might know. Sending and receiving parties apply a shrinking method to their bit sequences in order to obscure the eavesdropper ability to capture bit sequence [7, 8].

$$\text{Privacy bits} = L - M - s \quad \ldots (1)$$

L= bits result from RK, EE, and KR

M= expected values known by an eavesdropper

s = s is a constant chosen security parameter

## 4. RANDOMNESS TEST METHODS

A random number test can only try to detect distinct patterns or conditions which should exist with a small probability if a number truly random. Therefore, several test suites were developed and used for the testing task. Such tests are based like DIEHARD and NIST suite test [15,16].

This paper strives to provide a comprehensive analysis of the randomness of the quantum keys, also these suites were utilized for comparable task [14, 17]. TABLE 1 characterizes and summarizes the test suite properties for binary digits.

**Table 1. Tests suite proprieties**

| Random Number Test | Type | Min. (bits )strength |
|---|---|---|
| Frequency (Monobit) Test  NIST | Normal | minimum of 100 |
| Block Frequency  NIST, DIEHARD | $X^2$ | minimum of 100 |
| Runs Test | Normal | minimum of 100 |
| Longest Run of Ones in a Block  NIST | X2 | minimum bits for M 128 bits for M = 8 6,272 bits for M = 128 750,000 bits for M = 10,000 |
| Rank Test  NIST, DIEHARD | X2 | For M = Q = 32, each sequence to be tested should consist of a minimum of 38,912 bits. |
| Overlapping template  NIST, DIEHARD | X2 | assumed a sequence length of 1,000,000 bits |
| Non-overlapping Template Matching Test<br><br>NIST, DIEHARD | X2 | assumed a sequence length of 1,000,000 bits |
| Maurer's "Universal Statistical" Test<br><br>NIST | Normal | minimum of 387,840 bits |
| Linear Complexity Test | Normal | minimum of 1,000,000 bits |
| Serial Test<br><br>NIST, DIEHARD | X2 | m and n such that m < [log2 n] -2 |
| Count the 1s<br><br>DIEHARD | Normal | minimum of 100 |
| Parking lot test<br><br>DIEHARD | X2 | minimum of 100 |
| Maximums of sub-series<br><br>De Finetti's theorem | Normal | minimum of 100 |
| Uniform Distribution<br><br>NIST, DIEHARD | Normal | minimum of 100 |
| Poker test ( Kendall) | Normal | minimum of 100 |
| Extreme point<br><br>Alxnedra | $X^2$ | minimum of 100000 |

*Note: the minimum size of bits digit and type of computation methodology lead to produce the probability value (P->value), which is determined the randomness ratio in each binary distribution and generated.*

To compute the randomness for any binary distribution series, following steps must be follow:

- *Step1: Stat the null hypothesis (assume that the binary sequence is random).*
- *Step2: Compute a sequence test static with different test suite (based on the bit level)*
- *Step3: Calculate the P-value (always P-value$\in \{0,1\}$).*
- *Step4: Compare the P-value produced to $\alpha$- (usually $\alpha = 0.01$ for all randomness testers), if the P-value less than 0.01, then sequence digits generated are failing [15].*

The P-value measures the support for randomness hypothesis on the basis of a particular test

## 5. BB84-PROTOCOL IMPLEMENTATION AND SIMULATIOM

To examine the relation between the randomness and the QKD, The BB84 is simulated using a Core i3 (2.4 GHz) processor associated with 4GB of RAM. Figure 2 shows the variance of the obtained results when different numbers of photons are pumped.

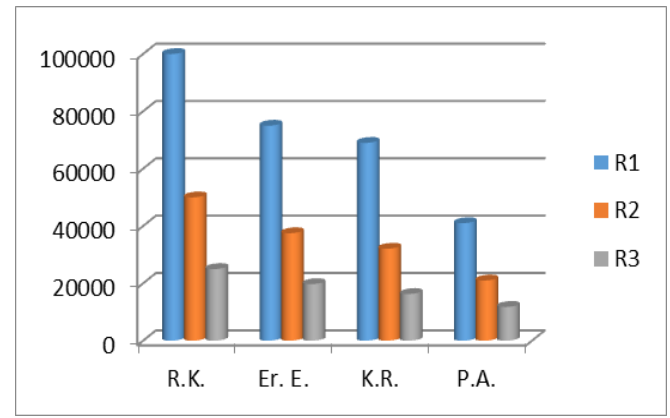

**Fig 2: qubits extraction based QKD**

For example, as shown in Figure2, at round 1, the raw key extraction holds 100000 qubits, which in turn are processed till the PA phase to succeed 41000 qubits. The obtained qubits diminution has occurred during the distillation processing, and the lost qubits are known as the authentication cost [19, 20].

## 6. TESTING AND ANALYSIS

This section discusses the testing phase for the qubits generated by the PA. As shown in TABLE 1, 14 random testing algorithms were implemented based this phase, these algorithms compute the coefficient P-value using non-deterministic hypothesis.

The random testing algorithms applied include 14 tests (see table 1), these algorithms compute a coefficient based on a non-deterministic hypothesis to obtain a measure of the randomness on bits are generated form QKDs depend on p-value, which is computing based on different mathematical operations . In this phase, 14 tests with three QKD rounds were implemented using the same qubits longest (100000-bits).

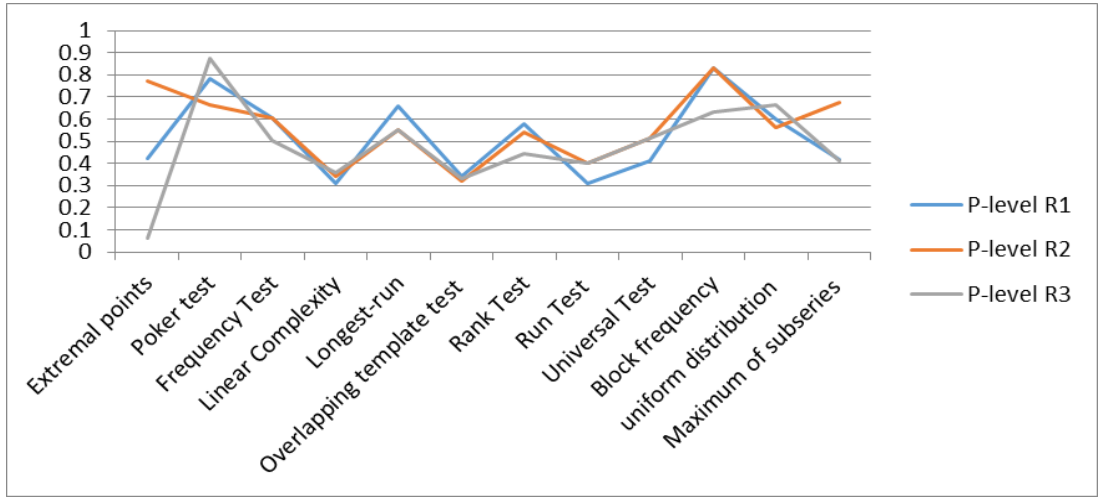

**Fig 3: P-values based on privacy bits and 12- tests suite, 5-series are included with each tester: with Uniform distribution, series ranging between (0.537804- 0.108736), 2 intervals, dimensions: 8; Maximum of subseries, series ranging between (0.537804- 0.108736), 10 intervals, sub-series length: 3; Frequency test, the nth partial sum = 518, $S\_n/n$ = 0.000518; Linear Complexity, M=500, N=2000; Longest Runs, N=100, M=1000, Chi^2 = 4.139447; Overlapping test, n=100000, m=9, M=1032, N=968 and lambda [(M-m+1)/2^m] = 2.00000; Rank test, Chi^2 = 1.096954;**

As shown in Figure (3), the results of the P-values periodically changed with the rounds contents, which verify the truly random generation for the qubits. As an example, if the frequency tests, the p-value is more than 0.521 and less than 0.687. Which in turn, based on NIST recommendations and others [15, 16, 18], indicates that the generated raw bits verifies the truly random characteristic.

Finally, the serial and cumulative random tests [13, 14] are implemented and the results are shown in Figure 4 and 5, despite the admission of the random generation, there is tiny un-equivalence between the P-values for the different bits (0-bits and 1-bits). The un-equivalence is caused by the influence of the noise and the transformation process.

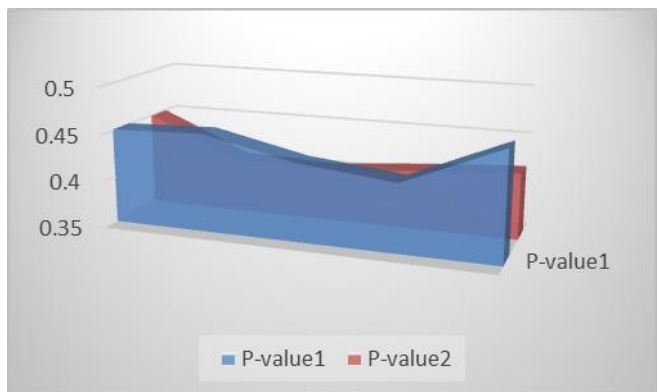

**Fig 4: P-valued based serial test per 3-rounds, m=5, n=10000**

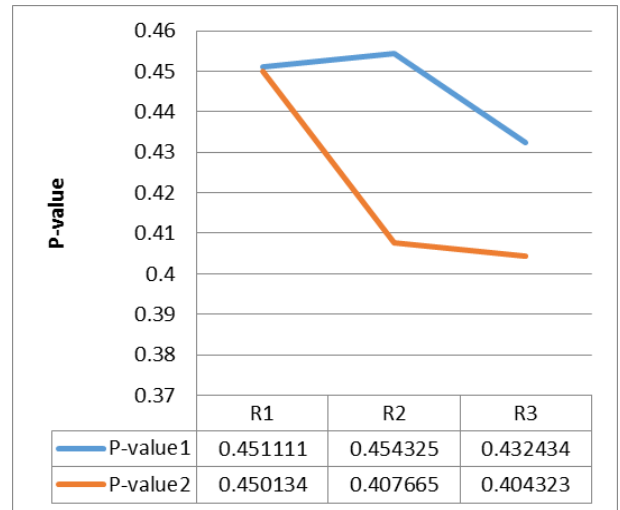

**Fig 5: P-valued based cumulative test per 3-rounds, m=5, n=100000**

## 7. CONCLUSIONS AND FUTURE WORKS

In this paper, the randomness of quantum key distribution system based BB84 protocol was investigated with suite of random number tests. Usually, the degree of randomness in QKD -bits are effected by the noise and the unrecognized the photon polarization. Moreover, a highly randomness can be gained using QKD, due to the entailed key distillation phases, such as ER, KR and PA. Finally, in this paper, the implementation reveals that all tests generate truly random bits except the serial test.

So, in the future work, the improvement of random generation considered, by developing a real hardware environment convenient with this simulator.

## 8. ACKNOWLEDGMENTS

The authors would like to thank the anonymous reviewers for their valuable comments and suggestions that improved the presentation of this paper. Moreover, many thanks to